\documentclass[twocolumn,english,aps,prl,showpacs]{revtex4}
\usepackage[T1]{fontenc}
\usepackage[latin9]{inputenc}
\usepackage{amstext}
\usepackage{graphicx}
\usepackage{esint}

\makeatletter
\@ifundefined{textcolor}{}
{%
 \definecolor{BLACK}{gray}{0}
 \definecolor{WHITE}{gray}{1}
 \definecolor{RED}{rgb}{1,0,0}
 \definecolor{GREEN}{rgb}{0,1,0}
 \definecolor{BLUE}{rgb}{0,0,1}
 \definecolor{CYAN}{cmyk}{1,0,0,0}
 \definecolor{MAGENTA}{cmyk}{0,1,0,0}
 \definecolor{YELLOW}{cmyk}{0,0,1,0}
}

\makeatother

\usepackage{babel}
\begin{document}

\title{Connection between heat diffusion and heat conduction in one-dimensional
systems}

\author{Shunda Chen}

\author{Yong Zhang}

\author{Jiao Wang}

\author{Hong Zhao}

\email{zhaoh@xmu.edu.cn}

\affiliation{Department of Physics and Institute of Theoretical Physics and Astrophysics,
Xiamen University, Xiamen 361005, Fujian, China}
\begin{abstract}
Heat and energy are conceptually different, but often are assumed to
be the same without justification. An effective method for investigating
diffusion properties in equilibrium systems is discussed. With this method, we demonstrate that for one-dimensional systems, using the indices of particles as the space variable , which has been accepted as a convention, may lead to misleading conclusions. We then show that though in one-dimensional systems there is no general connection between energy diffusion and heat conduction, however, a general connection between heat diffusion and heat conduction may exist. Relaxation behavior of local energy current fluctuations and that of local heat current fluctuations are also studied. We find that they are significantly different, though the global energy current equals the globe heat current.
\end{abstract}


\maketitle

\section{Introduction}

By definition, it is clear that heat and internal energy are conceptually
different. Internal energy is referred to as the total kinetic and
potential energy of a system, which is a function of the system state,
while heat is a quantity that characterizes a process. For one-dimensional
(1D) systems, combining the continuous equations of energy and mass,
i.e., $\frac{\partial e(x,t)}{\partial t}+\frac{\partial}{\partial x}{j}^{e}(x,t)=0$
and $\frac{\partial\rho(x,t)}{\partial t}+\frac{\partial}{\partial x}{p}(x,t)=0$,
one can obtain
\begin{equation}
\frac{\partial}{\partial t}[e(x,t)-\frac{(e+P)\rho(x,t)}{\rho}]+\frac{\partial}{\partial x}{j}^{q}(x,t)=0,
\end{equation}
 and thus introduce the heat density function as \cite{1,2,3,4} 
\begin{equation}
q(x,t)=e(x,t)-\frac{(e+p)\rho(x,t)}{\rho}.
\end{equation}
 Here $e(x,t)$, $\rho(x,t)$, ${p}(x,t)$, ${j}^{e}(x,t)$, and ${j}^{q}(x,t)$
represent, respectively, the density of energy, mass, momentum, energy
current, and heat current; $e$ ($\rho$) and $P$ represent, respectively,
the spatially averaged energy (mass) density and the internal pressure
of the system at the equilibrium state. The local heat current is
related to the local energy current as
\begin{equation}
{j}^{q}(x,t)={j}^{e}(x,t)-\frac{e+P}{\rho}{p}(x,t).
\end{equation}
 Therefore, the physical meaning of the change rate of $q(x,t)$ is
definite: it represents the divergence of the local heat current.
But, on the contrary, the value of $q(x,t)$ itself lacks definite
meanings. Usually $q(x,t)$ is negative, because $P>0$ and $\langle\rho(x,t)/\rho\rangle=1$.
Indeed, following Eq. (1), the heat density can be defined as $q(x,t)+c$
with an arbitrary constant $c$. Energy contains two parts, corresponding
to regular and irregular motions, respectively, but heat is always
related to thermal processes that feature random motions.

In spite of this fact, in some previous studies energy and heat are
not distinguished. An example is in the study of the relation between
anomalous diffusion and transport properties in low-dimensional (one-
and two-dimensional) systems. It is well known that in bulk (three-dimensional)
materials the thermal conductivity $\kappa$ and the heat diffusion
coefficient $D$ can be generally related by $\kappa=\rho c_{P}D$,
where $c_{P}$ is the constant pressure specific heat. But this relation
does not hold in low-dimensional systems. In recent decades, stimulated
by the rapid progress in nanoscience \cite{5,6}, transport properties
of low-dimensional systems has attracted intensive investigations
\cite{7,8,9,10,11,12,13}. 
It has been found that in general diffusion and transport are abnormal
in low-dimensional systems. In particular, in 1D momentum conserving
systems, the heat conductivity diverges with the system size $L$
as $\kappa\sim L^{\alpha}$ and the heat diffusion coefficient diverges
with time as $D\sim t^{\beta-1}$. ($\alpha$ and $\beta$ are constants.)
In some studies \cite{14,15} 
the heat diffusion behavior has been assumed, implicitly, to be the
same as the energy diffusion behavior, and the exponent $\beta$ is
calculated by tracing \textit{energy} diffusion instead.

It has been conjectured that there exists a general relation between
exponents $\alpha$ and $\beta$. Two formulae, $\alpha=2-2/\beta$
by Li and Wang \cite{16} 
and $\alpha=\beta-1$ by Denisov et al. \cite{17}, 
have been proposed. It is worth noting that in these studies \cite{16,17}
the authors did not distinguish heat diffusion and energy diffusion
and treated them identically. Though this is legal for the specific
example systems they studied where energy and heat are coincidentally
the same, we emphasize that in principle $\beta$ involved in these
two formulae should be the exponent that characterizes \textit{heat}
diffusion, rather than that of energy diffusion. This is particularly
important for clarifying which one of these two formula is correct,
which has been a focused issue recently. Indeed, in our very recent
study \cite{18} 
it has been shown that the diffusion behaviors of energy and heat
can be qualitatively different, suggesting that there should be no
general relation between energy diffusion and heat conduction, but
a general relation between \textit{heat diffusion} and \textit{heat
conduction} may exist and can be established.

This paper is an effort to verify this conjecture. We shall first
ascertain a correct way for calculating the exponent $\beta$, so
as to put our studies on a solid basis. So far there are two classes
of methods for probing diffusion processes, i.e., the equilibrium
\cite{15,18,19}
and non-equilibrium \cite{14,20,21,22,23,24} 
methods. With both methods, the probability density function (PDF)
of local fluctuations of interested physical quantities are calculated.
The space variable of the PDF should correctly give the positions
of local fluctuations, but in practice the indices of particles are
often taken as the space variable instead to facilitate the simulations
\cite{14,15}. 
In doing so, the underlying philosophy is that the index of a particle
is equivalent to its position in 1D systems, as the particle simply
moves around its equilibrium position. But as we shall demonstrate
in the following, this is not the case: Using the index variable may
result in not only quantitative but also qualitative deviations, which
may be responsible for the confusing results of $\beta$ reported
previously in the literatures. By taking the correct space variable
and the equilibrium method (which has been shown to be more efficient
and accurate), we calculate the exponent $\beta$ of heat diffusion
with high precision in a 1D hard-point gas model \cite{14,25,26,27}
By comparing it with the values of the exponent $\alpha$ obtained
in previous studies \cite{27,28,29,30} 
we shall show that the anomalous heat diffusion and heat conduction
can be accurately connected by the formula $\alpha=2-2/\beta$.

In addition, we shall also discuss the behaviors of the local heat
current and the local energy current. By properly setting the coordinate
system to guarantee the system has a vanishing total momentum, we
find although the total heat current is always equal to the total
energy current, the relaxation behaviors of local currents of energy
and heat can be remarkably different.

The rest of this paper is organized as follows: The model to be studied
will be described in the next section, and the methods for probing
energy and heat diffusion will be detailed in Sec. 3. The main results
will be provided and discussed in Sec. 4, followed by a brief summary
in the last section.

\section{Models}

We consider two paradigmatic 1D models extensively employed for studying
transport properties of low-dimensional systems. Each model is composed
of $N$ point particles arranged in order. We denote by $m_{k}$,
$x_{k}$, $v_{k}$, and $p_{k}$, respectively, the mass, the position,
the velocity, and the momentum of the $k$ particle.

The first model is a 1D hard-point gas \cite{14,25,26,27} 
with alternative mass $m_{o}$ for odd-numbered particles and $m_{e}$
for even-numbered particles. We set $m_{o}=1$ and $m_{e}=3$, the
same as in Ref. \cite{14}, 
for the sake of comparison. The particles travel freely except for
elastic collisions with their nearest neighbors. After a collision
between the $k$th particle and the $(k+1)$th particle, their velocities
change to
\begin{equation}
v_{k}^{\prime}=\frac{m_{k}-m_{k+1}}{m_{k}+m_{k+1}}v_{k}+\frac{2m_{k+1}}{m_{k}+m_{k+1}}v_{k+1},
\end{equation}

\begin{equation}
v_{k+1}^{\prime}=\frac{2m_{k}}{m_{k}+m_{k+1}}v_{k}-\frac{m_{k}-m_{k+1}}{m_{k}+m_{k+1}}v_{k+1}.
\end{equation}
 Another model is a 1D lattice; i.e., the well known Fermi-Pasta-Ulam
(FPU) model defined by the Hamiltonian
\begin{equation}
H=\sum_{k}\frac{p_{k}^{2}}{2m_{k}}+\frac{1}{2}(x_{k}-x_{k-1}-1)^{2}+\frac{1}{4}(x_{k}-x_{k-1}-1)^{4},
\end{equation}
 where the masses of all particles are set to be unity.

In our simulations the periodic boundary condition is applied and
the system size $L$ is set to be the same as the particle number
$N$, so that the averaged particle number density is unity. The local
temperature is defined as $T_{k}\equiv\frac{\langle p_{k}^{2}\rangle}{k_{B}m_{k}}$,
where $k_{B}$ (set to be unity) is the Boltzmann constant and $\langle\cdot\rangle$
stands for the ensemble average. For both models the average energy
per particle is fixed to be unity, corresponding to a system temperature
$T=2$ in the gas model and $T\approx1.2$ in the FPU model.

\section{Methods}

In principle, one can probe the diffusion behavior directly by adding
an external perturbation to the equilibrium system and observing its
ensuing relaxation process \cite{21,22,23}. 
This method requires demanding computing resource, so that a satisfactory
precision is usually hard to reach \cite{19}. 
A more effective method \cite{15,18} 
is instead to study the properly rescaled spatiotemporal correlation
functions of fluctuations in the equilibrium state. The basic idea
of this method is detailed in the following.

Let $A(x,t)$ represents the density distribution function of a physical
quantity ${\cal A}$. In numerical simulations, in order to calculate
the spatiotemporal correlation function of fluctuations of ${\cal A}$,
we have to discretize the space variable first. For this aim, we divide
the space occupied by the system into $N_{b}=L/b$ bins of equal size
of $b$. The total quantity of ${\cal A}$ in the $k$th bin, denoted
by $A_{k}(t)$, is defined as $A_{k}(t)\equiv\int_{x\in k\mathrm{th\, bin}}A(x,t)dx$.
As such $A_{k}(t)/b$ gives the coarse-grained density of ${\cal A}$
in the $k$th bin. The fluctuations of ${\cal A}$ are therefore captured
by $\Delta A_{k}(t)\equiv A_{k}(t)-\langle A\rangle$, where $\langle A\rangle$
represents the ensemble average of $A_{k}(t)$. The positions of the
bin centers can then be used as the coarse-grained space variable.

\begin{figure*}
\vskip-.2cm 
\includegraphics[scale=0.5]{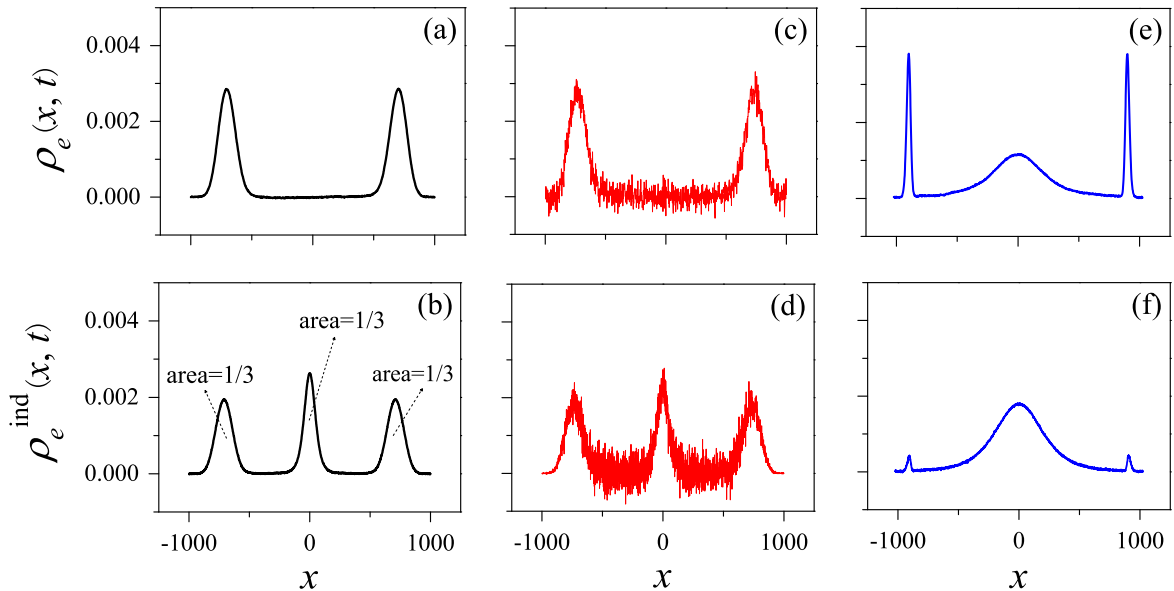} \vskip-.8cm \caption{The probability density distribution function of energy obtained by
using our correlation function method and the perturbation method
with the space variable being, respectively, the coarse-grained space
variable and the particle indices. (a)-(d) are for the gas model at
$t=400$, obtained by (a)-(b) the correlation function method and
(c)-(d) the perturbation method. (e)-(f) are for the FPU model at
$t=600$ by using the correlation function method.}
\end{figure*}


For a conserved physical quantity ${\cal A}$, it has been derived
in Ref. \cite{18} 
that the PDF corresponding to a local fluctuation initially located
in the $k$th bin, which is specified by $\Delta A_{k}(0)$, can be
calculated as
\begin{equation}
\rho_{A}(\Delta x_{k,l},t)=\frac{\langle\Delta A_{l}(t)\Delta A_{k}(0)\rangle}{\langle\Delta A_{k}(0)\Delta A_{k}(0)\rangle}+\frac{1}{N_{b}-1}
\end{equation}
 if the microcanonical ensemble is considered, and
\begin{equation}
\rho_{A}(\Delta x_{k,l},t)=\frac{\langle\Delta A_{l}(t)\Delta A_{k}(0)\rangle}{\langle\Delta A_{k}(0)\Delta A_{k}(0)\rangle}
\end{equation}
 if the canonical ensemble is considered. Here $\Delta x_{k,l}$ denotes
the displacement from the $k$th bin to the $l$th bin, i.e., $\Delta x_{k,l}\equiv(l-k)b$.
For the sake of convenience, in the following we shall use $x$ to
denote $\Delta x_{k,l}$ without confusion. The spatiotemporal correlation
function defined above gives the causal relation between a local fluctuation
and the effects it induces at another position and at a later time,
thus it is in essence equivalent to the probability density function
that describes the diffusion process of the fluctuation. In order
to facilitate numerical simulations, we suggest considering the microcanonical
ensemble where all systems are isolated from the environment. Hence
one does not have to simulate the environment, which saves greatly
the simulation time.

In previous studies, e.g., \cite{14,15,22,23,24}, 
the authors constantly use the indices of particles to represent the
space variable. In particular, the value of ${\cal A}$ of the $k$th
particle, denoted by $A_{k}^{{\rm ind}}(t)$, is adopted to represent
the density distribution of ${\cal A}$ at the position of $kL/N$,
and $\langle\Delta A_{l}^{{\rm ind}}(t)\Delta A_{k}^{\text{i}nd}(0)\rangle$
is assumed to represent the correlation between two positions with
a distance of $x=(l-k)L/N$ and a time delay of $t$. In the following,
we shall refer to this ``coordinate\textquotedbl{} as the index variable.
Although the index represents the \textit{mean} position of a particle
in the equilibrium state, it by no means gives the position of the
particle at instant times that is crucial for correctly calculating
the spatiotemporal correlation functions. For this reason, Dhar once
questioned the effectiveness of the index variable because it may
result in large position fluctuations \cite{31}. 
We find that it is even worse: The deviations caused by using the
index variable is not only quantitative, but also qualitative.

\begin{figure*}
\vskip-.4cm \hskip-.3cm \includegraphics[scale=0.48]{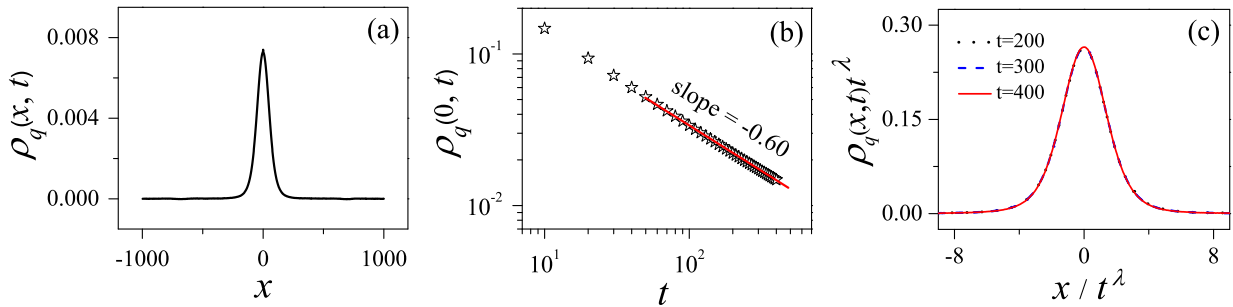} \vskip-.8cm
\caption{The simulation results of the spatiotemporal correlation function
of heat fluctuations, $\rho_{q}(x,t)$, for the 1D gas model. (a)
presents a snapshot of $\rho_{q}(x,t)$ at time $t=400$, and (b)
shows the time dependence of the height of the center peak of $\rho_{q}(x,t)$.
The red solid line in (b) is the best linear fitting of the data for
revealing their asymptotic characteristics, suggesting that $\rho_{q}(0,t)\sim t^{-0.60}$.
In (c) $\rho_{q}(x,t)t^{\lambda}$ versus $x/t^{\lambda}$ at three
different times are compared with the rescaling factor $\lambda=0.60$
obtained via best linear fitting in (b). The fact that three curves
overlap perfectly verifies the scaling property of $\rho_{q}(x,t)$.}
\end{figure*}


Taking the energy fluctuations as an example, we show that indeed
the index variable may lead to qualitatively wrong results. We denote
the spatiotemporal correlation function obtained by using the coarse-grained
space variable and the index variable as $\rho_{e}(x,t)$ and $\rho_{e}^{{\rm ind}}(x,t)$,
respectively. The 1D gas model is considered first. To prepare an
equilibrium gas, the system is efficiently simulated for a sufficient
long time by using the event-driven algorithm that employs the heap
data structure to identify the collision times \cite{27}. 
Then $\rho_{e}(x,t)$ and $\rho_{e}^{{\rm ind}}(x,t)$ are calculated
with $N=4000$ and $b=1$. Figure 1(a)-(b) show the results. One can
see that they are remarkably different: With the coarse-grained space
variable, the spatiotemporal correlation function has two peaks, while
with the index variable it has three peaks. The two peaks of $\rho_{e}(x,t)$
move outwards with a constant speed $v=1.75$, which can be shown
easily to be the sound speed \cite{18}. 
The two side peaks of $\rho_{e}^{{\rm ind}}(x,t)$ move outwards with
the same speed. The center peak of $\rho_{e}^{{\rm ind}}(x,t)$ does
not move but broadens as $\Delta w\sim t^{0.67}$, where $\Delta w$
represents its half-height width. It is tempting to think that the
two side peaks represent the sound mode and the center peak represents
the heat mode, as in the case of the mass fluctuations that gives
the dynamic structure factor of the system \cite{4}. 
But we find the ratio of the area of the center peak to that of the
two side peaks equals $1/2$, while it should equal $2$, i.e., the
Landau-Placzek ratio \cite{32,33} 
of an ideal gas, if it characterizes the dynamical structural factor
\cite{4}. 
As will be discussed in the next section, the decaying behavior of
the center peak is also different from the heat mode. Therefore, $\rho_{e}^{{\rm ind}}(x,t)$
fails to capture the properties of the heat mode. It is worth noting
that the three peak structure has also been reported in other studies
of the gas model by using the index variable \cite{14,23,24}. 

As mentioned above, diffusion properties can also be investigated
directly by observing the evolution of an added perturbation to the
system. We find that by applying this method, one can obtain the same
results. {[}See Fig. 1(c)-(d).{]} In addition, we find that in the
1D FPU model {[}see Fig. 1(e)-(f){]}, the qualitative difference between
$\rho_{e}(x,t)$ and $\rho_{e}^{{\rm ind}}(x,t)$ is also obvious
\cite{18}. These results suggest clearly that the position of a particle
at instant times cannot be approximated by its equilibrium position
in order to correctly calculate spatiotemporal correlation functions.

\section{Results and Discussions}

In Fig. 1(a), it is clearly shown that the energy fluctuations transport
ballistically in the 1D gas model, implying that the mean square displacement
of the transported energy increases in time as $\langle x^{2}(t)\rangle\sim t^{2}$.
On the other hand, the heat conduction properties of this model have
been extensively studied in the literatures \cite{25,26,27}. 
It has been found that the heat conductivity $\kappa$ diverges with
the system size $L$ as $\kappa\sim L^{\alpha}$ with $\alpha=1/3$
\cite{27,28,29,30}, 
suggesting that heat diffuses in a supperdiffusive manner rather than
ballistically. Therefore, energy diffusion and heat conduction do
not fall into the same anomalous class. We then calculate the PDF
of heat fluctuations, i.e., $\rho_{q}(x,t)$, following Eq. (2) and
present the results in Fig. 2. It can be seen that the profile of
$\rho_{q}(x,t)$ is completely different from that of energy fluctuations,
$\rho_{e}(x,t)$ {[}see Fig. 1(a){]}; the former has only one single
peak. This outstanding feature has been reported in ref. \cite{18},
but here we perform the simulation with a larger system size of $N=4096$
which allows us to measure the decaying rate of $\rho_{q}(x,t)$ more
accurate. As presented in Fig. 2(b), we obtain that the height of
the peak of $\rho_{q}(x,t)$ goes as $h=\rho_{q}(0,t)\sim t^{-\lambda}$
with $\lambda=0.60$. As heat is a conserved quantity, the peak of
$\rho_{q}(x,t)$ must keep its area unchange, and as a consequence
its half-height width should broaden as $\Delta w\sim t^{-0.60}$.
As such $\rho_{q}(x,t)$ at different time may be rescaled onto a
common function. This is confirmed by our study presented in Fig.
2(c): $\rho_{q}(x,t)$ is indeed invariant upon rescaling $x\rightarrow t^{\lambda}x$
so that $t^{\lambda}\rho_{q}(x,t)=t_{0}^{\lambda}\rho_{q}(x_{0},t_{0})$
for $x=({t}/{t_{0}})^{\lambda}x_{0}$ with scaling factor $\lambda=0.60$.

\begin{figure*}
\hskip-.1cm \includegraphics[scale=0.48]{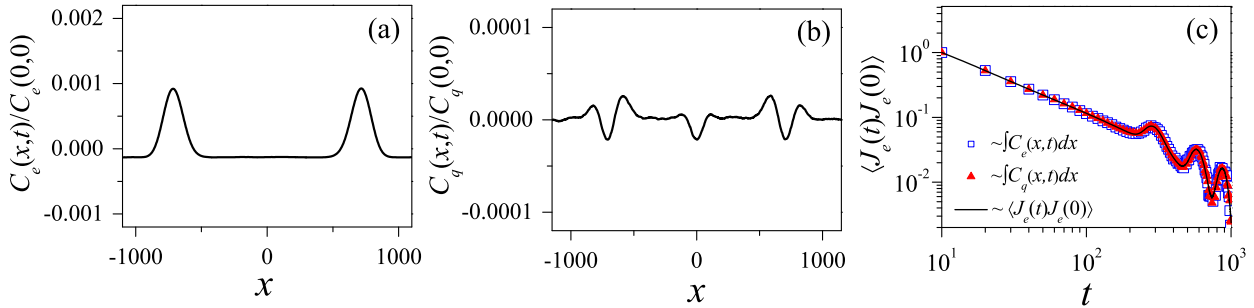} \vskip-.8cm \caption{The spatiotemporal correlation function of local energy current $C_{e}(x,t)$
(a) and of local heat current $C_{q}(x,t)$ (b) at time $t=400$ for
the 1D gas model $N$=3000. In (c), the autocorrelation function of
the globe energy current $\langle J_{e}(t)J_{e}(0)\rangle$ (black
solid line) is compared with $\int C_{e}(x,t)dx$ (blue open squares)
and $\int C_{q}(x,t)dx$ (red solid triangles). The three sets of
data fall onto the same curve upon proper shifts in vertical direction.
$N$=512.}
\end{figure*}


This result confirms for the first time that the relaxation of the
heat mode follows the scaling law with $\lambda=3/5$ as predicted
by the hydrodynamic mode-coupling theory \cite{34}. 
This scaling property implies that $\rho_{q}(x,t)$ relaxes as $\langle x^{2}(t)\rangle=\langle x_{0}^{2}(t_{0})\rangle(\frac{t}{t_{0}})^{2\lambda}$;
i.e., heat fluctuations diffuse in a power law $\langle x^{2}(t)\rangle\sim t^{\beta}$
with the diffusion exponent $\beta=2\lambda$ \cite{18}. 
We obtain $\beta=1.20$ accordingly, suggesting that heat diffusion
is superdiffusive, in clear contrast with energy diffusion which is
ballistic. More important and interesting, by combining $\beta=1.20$
obtained here and $\alpha=1/3$ obtained in previous analytical and
numerical studies \cite{27,28,29,30}, 
we find they follow perfectly the general formula proposed by Li and
Wang \cite{16} 
that trying to connect energy diffusion (should be heat diffusion
instead) and heat conduction in 1D systems.

We would like to point out that the center peak of $\rho_{e}^{{\rm ind}}(x,t)$
{[}see Fig. 1(a){]} is also invariant upon the rescaling but with
$\lambda=0.67$ instead, which implies $\beta=1.34$. One may notice
that the formula $\alpha=\beta-1$ correctly describes the relation
of energy diffusion and heat conduction in this case, as pointed in
Ref. \cite{14,15}. 
We can therefore conclude that this is a consequence by improperly
replacing the space variable with the particle indices.

Finally, we study the relaxation behavior of local energy current
and local heat current. We set the total momentum of the system to
be zero. As having been pointed out in Ref. \cite{18}, 
the global energy current always equals the globe heat current, but
this fact does not imply the properties of local heat current and
local energy current are also identical. Figure 3(a)-(b) show the
spatiotemporal correlation functions of local energy current $C_{e}(x,t)=\langle j_{l}^{e}(t)j_{k}^{e}(0)\rangle$
and that of local heat current $C_{q}(x,t)=\langle j_{l}^{q}(t)j_{k}^{q}(0)\rangle$.
Here $x=(l-k)L/N$, ( the size of a bin is $b=L/N$ ), the local energy
current is defined as $j_{l}^{e}(t)=\sum_{k:x_{k}\in l{\rm th~bin}}m_{k}v_{k}^{3}/2$,
and the local heat current is obtained by substituting $j_{l}^{e}(t)$
into Eq. (3). It can be seen that $C_{e}(x,t)$ and $C_{q}(x,t)$
have remarkably different features: The former has a global negative
bias, and its two peaks moving oppositely at the sound speed. The
latter is more complicated. There are two pulses which look like a
negative Mexican hat wavelet and move outwards at the sound speed,
but there is no global bias. Instead, there is a dip at the origin.
Therefore, the properties of $C_{q}(x,t)$ can not be probed by studying
$C_{e}(x,t)$, and vice versa.

The globe energy current $J_{e}(t)=\sum_{k}j_{k}^{e}(t)$. One has
$\langle J_{e}(t)J_{e}(0)\rangle=\langle\sum_{k}j_{k}^{e}(0)\sum_{l}j_{l}^{e}(t)\rangle\propto\langle j_{k}^{e}(0)\sum_{l}j_{l}^{e}(t)\rangle$
for any index $k$ because of the homogeneity of the system. In other
words, $\langle J_{e}(t)J_{e}(0)\rangle\propto\int C_{e}(x,t)dx$.
Because $J_{e}(t)=J_{q}(t)$ as a result of the null total momentum,
we have $\int C_{e}(x,t)dx\propto\int C_{q}(x,t)dx$. This result
implies that though the relaxation behaviors of local heat current
and local energy current are different, their integrals decay in the
same way, which is also confirmed by direct simulation results shown
in Fig. 3(c).

\section{Summary}

We demonstrate that in 1D systems, using the particle indices as the
space variable may result in qualitative deviations in probing the
diffusion properties, and hence should be abandoned. Instead, the
coarse-grained space variable is a correct and practical choice. By
taking advantage of it, we have verified that in the 1D gas model,
heat diffusion and heat conduction, rather than energy diffusion and
heat conduction, can be connected by the formula $\alpha=2-2/\beta$
\cite{16} 
accurately, rather than by that proposed in \cite{17}. 
Our analysis has also shown that 
energy diffusion and heat conduction follows $\alpha=\beta-1$ as
observed in Ref. \cite{14,15} 
may be a misunderstanding caused by misusing the particle indices
as the space variable. We emphasize that the position of a particle
at instant times cannot be approximated by its equilibrium position
in probing spatiotemporal correlation functions of a system.

In addition, in the case of null total momentum, the globe energy
current and the globe heat current are found to be the same. But local
energy currents are different from local heat currents. The relaxation
behavior of the former is significantly different from that of the
latter as well. We conclude that in general, the relaxation and transport
properties of heat can not be identified with those of energy.
\begin{acknowledgments}
This work is supported by the NNSF (Grants No. 10925525, No. 11275159,
and No. 10805036) and SRFDP (Grant No. 20100121110021) of China.


\end{acknowledgments}

\end{document}